\documentclass[11pt,twoside]{article}
\usepackage{./asp2014}

\aspSuppressVolSlug
\resetcounters

\begin{document}

\title{Stellar archaeology with Gaia: the Galactic white dwarf
population}
\author{
Boris G\"ansicke$^1$,
Pier-Emmanuel Tremblay$^{2,1}$,
Martin Barstow$^3$,
Giuseppe Bono$^4$,
Matt Burleigh$^3$,
Sarah Casewell$^3$,
Vik Dhillon$^5$,
Jay Farihi$^6$,
Enrique Garcia-Berro$^{7,8}$,
Stephan Geier$^9$,
Nicola Gentile-Fusillo$^1$,
JJ Hermes$^1$,
Mark Hollands$^1$,
Alina Istrate$^{10}$,
Stefan Jordan$^{11}$,
Christian Knigge$^{12}$,
Christopher Manser$^1$,
Tom Marsh$^1$,
Gijs Nelemans$^{13,14}$,
Anna Pala$^1$,
Roberto Raddi$^1$,
Thomas Tauris$^{10}$,
Odette Toloza$^1$,
Dimitri Veras$^1$,
Klaus Werner$^{15}$,
David Wilson$^1$
\affil{$^1$University of Warwick; $^2$STScI, USA; $^3$University of
  Leicester, UK; $^4$Universit\'a di Roma, Italy; $^5$University of
  Sheffield, UK; $^6$University College London, UK; $^7$Universitat
  Polit\`ecnica de Catalunya, Spain; $^8$ Institut d'Estudis Espacials
  de Catalunya, Spain; $^9$European Southern Observatory, Germany;
  $^{10}$Universit\"at Bonn, Germany; $^{11}$Universit\"at Heidelberg,
  Germany; $^{12}$University of Southampton, UK; $^{13}$Radboud
  University Nijmegen, Netherlands; $^{14}$KU Leuven, Belgium;
  $^{15}$Eberhard Karls Universit\"at T\"ubingen, Germany }}


\begin{abstract}
Gaia will identify several $10^5$ white dwarfs, most of which will be
in the solar neighborhood at distances of a few hundred
parsecs. Ground-based optical follow-up spectroscopy of this sample of
stellar remnants is essential to unlock the enormous scientific
potential it holds for our understanding of stellar evolution, and the
Galactic formation history of both stars and planets.
\end{abstract}

\section{White dwarfs as tracers of the Galactic star formation history}

The local white dwarf population preserves detailed information on the
Galactic star formation history. Nearly 95\% of all stars will end
their lives as white dwarfs, and most $\gtrsim1.2M_\odot$ stars ever formed
are now stellar remnants. White dwarf cooling is well understood in
terms of the underlying physics \citep{fontaineetal01-1}, and has been
used to estimate the age of the Galactic disc
(e.g. \citealt{oswaltetal96-1}), individual open clusters
(e.g. \citealt{garcia-berroetal10-1}), and the halo
\citep{kalirai12-1}.

Broader application of this method has been prevented so far by the
small and incomplete samples of known white dwarfs. Because of their
small radii, white dwarfs are intrinsically faint, requiring
moderately large telescope apertures for their study.  Consequently
the currently available luminosity functions contain at best a few
thousand stars \citep{harrisetal06-1, degennaroetal08-1}, based on
white dwarfs identified by SDSS. However, incompleteness and selection
biases severely limit the constraints that can be drawn from these
studies. \citet{tremblayetal14-1} demonstrated the potential of a
complete and well-characterised white dwarf sample to derive the local
star formation history and initial mass function, using only the 117
white dwarfs known within 20\,pc of the Sun, though obviously limited
by small number statistics.

\begin{figure}[!t]
\plotone{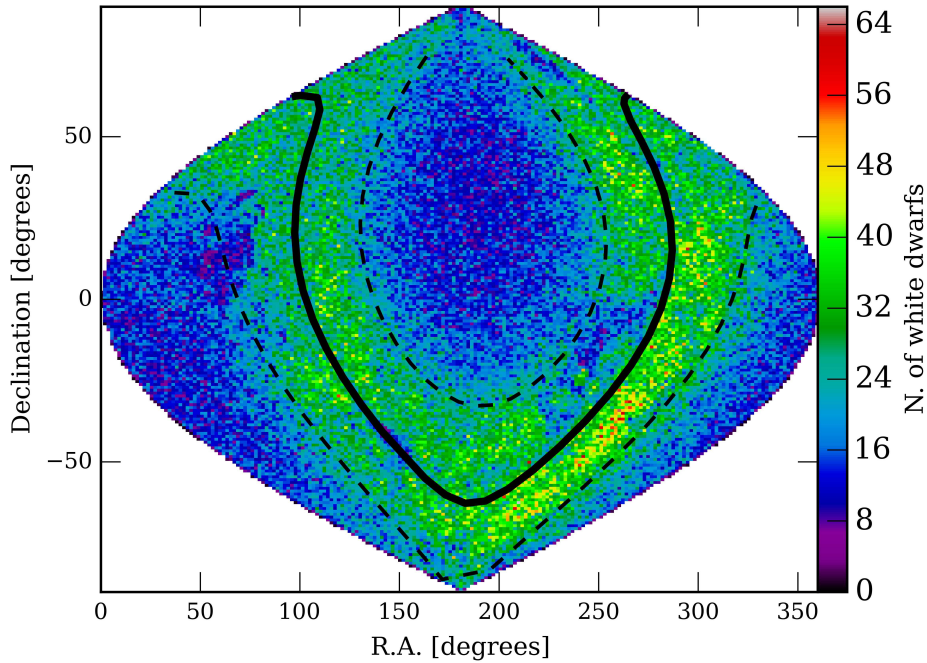}
\caption{\label{f:gaia_wds} Based on the GUMS-10 simulation (Robin et
  al. 2012), Gaia will detect $\simeq575\,000$ white dwarfs with
  $G<20$. The surface density of white dwarfs varies (shown here in
  deg$^{-2}$) between $\simeq5\,\mathrm{deg^{-2}}$ near the Galactic
  poles and $\simeq45\,\mathrm{deg^{-2}}$ in the Galactic plane (solid
  line, the dashed lines are $|b|=\pm30\deg$). The increase at low
  $|b|$ is largely due to young, bright, and more distant white
  dwarfs.}
\end{figure}

Combining parallax, apparent magnitude, proper motion and BP/RP
colour, Gaia will overcome all the traditional limitations in the
discovery of white dwarfs. The $\simeq5\times10^5$ white dwarfs that
Gaia will identify (Fig.\,\ref{f:gaia_wds}) will be 100\,\% complete
within $\simeq50$\,pc, and $\simeq50$\,\% out to 300\,pc
\citep{jordan07-1, carrascoetal14-1}. The sample of Gaia white dwarfs
will be sufficiently large to reconstruct the stellar formation
history and initial mass function for the thin/thick disc and halo
separately and assess the ages of these three components individually.

\begin{figure}[!t]
\plotone{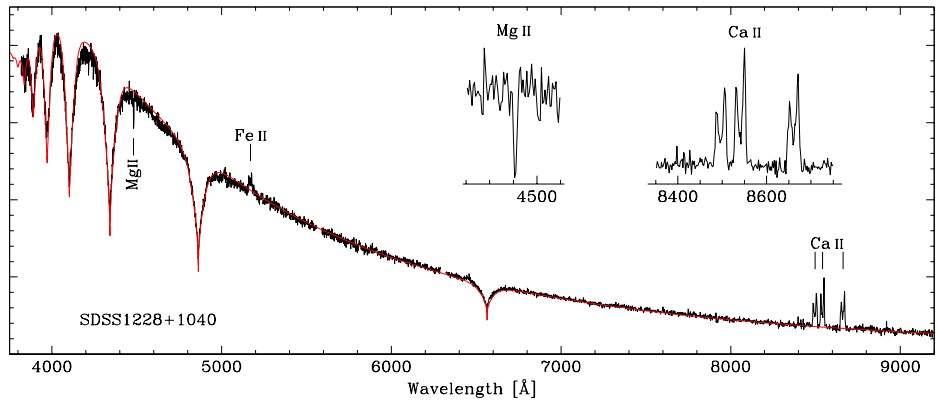}
\caption{\label{f:sdss1228} Most white dwarfs have hydrogen-dominated
  atmospheres (black). $T_\mathrm{eff}$ and $\log g$ are measured from
  fitting the Stark-broadened Balmer lines (red). Ground-based
  spectroscopy of the Gaia white dwarfs is essential, as the BP/RP
  spectra have insufficient spectral resolution. SDSS demonstrated the
  potential for serendipitous discoveries, such as this white dwarf
  accreting metal-rich planetary debris \citep{gaensickeetal06-3}.}
\end{figure}

However, while Gaia will identify the entire local white dwarf
population, it will not provide the data that is necessary to
determine accurate masses and temperatures, which are the key
parameters for measuring the white dwarf cooling ages. The spectral
resolution of the BP/RP spectra is too low to apply the standard
method for measuring temperatures and surface gravities from the
Stark-broadened Balmer line profiles (Fig.\,\ref{f:sdss1228}; e.g.
\citealt{bergeronetal92-1}). This implies that ground-based follow-up
is essential to unlock the diagnostic potential that the Gaia white
dwarfs have for investigating the Galactic star formation history. To
break the degeneracy between temperature and surface gravity requires
spectroscopy covering the higher Balmer lines down to $\simeq380$\,nm
\citep[e.g.][]{kepleretal06-1}, which are most sensitive to surface
gravity, and hence mass. Several of the forthcoming wide-area MOS
instruments are perfectly suited for this goal, including DESI, WEAVE,
and 4MOST.

Gaia will also not obtain radial velocities of white dwarfs, as they
typically have no spectral features in the RVS wavelength range, and
are too faint for RVS spectroscopy anyway. To establish the Galactic
orbits, and hence thin/thick disc or halo membership, radial
velocities of the Gaia white dwarfs have to be obtained from the
ground. This is most efficiently done by intermediate-resolution
($R\simeq5000$) spectroscopy of the sharp NLTE core in H$\alpha$
\citep{paulietal06-1, falconetal10-1}.  

\section{Additional science goals}

\textit{The initial-to-final mass (IFM) relation.} The physics of mass
loss on the RGB/AGB is still poorly understood, and lacks stringent
observational constraints. Yet, it is of crucial importance for the
Galactic lifecycle of matter, and for the chemical evolution of
galaxies \citep[e.g.][]{marastonetal98-1}. Traditionally, the IFM
relation has been investigated using white dwarfs in open clusters
\citep{casewelletal09-1}, and Gaia will significantly add to the known
population of white dwarfs in open clusters. Gaia, coupled with
ground-based spectroscopy will also fully enable the potential of wide
binaries to constrain the low-mass end of the IFM relation
\citep{catalanetal08-2, girvenetal10-1}.

\textit{The mass-radius (M-R) relation.}  Despite its fundamental
importance, from predicting white dwarf masses to understanding the
nature of SN\,Ia, this relation is very poorly constrained by
observations \citep{holbergetal12-1}. Spectroscopic observations of
white dwarfs cannot directly determine both mass and radius, and
therefore parallaxes are needed to independently determine the radius.
Currently only a handful of white dwarfs have parallaxes accurate to
better than 5\,\%, but Gaia will improve this situation dramatically,
both in terms of raw numbers of white dwarf parallaxes, as well as
their accuracy. Spectroscopic follow-up covering the higher Balmer
lines is essential to reach a few percent precision on both mass and
radius. The internal stratification of white dwarfs can then be
constrained by comparing the observed M-R relation with theoretical
relations assuming different interior compositions (3--5\,\% effect on
the M-R relation). Accurate knowledge of the core compositions will
also vastly improve the derived cooling ages, which are essential for
the Galactic archeology applications.

\textit{Rare white dwarfs.} SDSS has thoroughly demonstrated the
\textit{huge} discovery potential of a large spectroscopic survey of
white dwarfs. Follow-up of the Gaia white dwarfs will be essential to
improve our understanding of rare evolution channels such as
C/O-atmosphere white dwarfs, that may be remnants of S-AGB stars
(e.g. \citealt{dufouretal07-1, gaensickeetal10-1}), strongly magnetic
white dwarfs that can be used as laboratories for physics under
extreme conditions \citep{kuelebietal09-1}, and dynamically active
evolved planetary systems (e.g. \citealt{gaensickeetal06-3}, see
Fig.\,\ref{f:sdss1228}). Throughout the last decade, it has become
clear that white dwarfs serve not only as tracers of stellar evolution
and star formation, but provide equally important information on the
formation, structure, and evolution of planetary systems. Many white
dwarfs ($>30$\,\%) show signatures of planetary systems in the form of
metal-pollution of their otherwise gravitationally settled pure-H/He
atmospheres (e.g. \citealt{zuckermanetal10-1}). These metals, accreted
from disrupted planetary bodies, provide the only method to directly
measure the bulk composition of extra-solar planetary systems,
including the detection of water-rich planetesimals
(e.g. \citealt{farihietal13-2}). Gaia follow-up spectroscopy will be
essential to increase the known sample of metal-polluted white
dwarfs. The strongest tracer of planetary debris is the Ca\,H/K
doublet, again requiring blue coverage down to $\simeq380$\,nm at
intermediate-resolution to detect the most polluted objects.

\textit{SN\,Ia progenitors.} Although double-degenerate white dwarf
binaries are one of the likely populations to produce SN\,Ia, the
current census of these stars is utterly inadequate to test population
models and predictions of SN\,Ia rates and delay time distributions
\citep{maozetal14-1}. The largest dedicated search for
double-degenerates covered only a few hundred stars
\citep{napiwotzkietal01-1}, and found a close binary fraction of
$\sim5$\,\%. Time-resolved spectroscopy of the Gaia white dwarfs using
DESI/WEAVE/4MOST sub-spectra has the potential to identify 1000s of
close double-degenerates, spanning the entire parameter space in mass,
mass ratio, and orbital period.

\section{Survey requirements}
The key goal is to obtain low resolution ($R\simeq5000$) spectroscopy
of the entire local white dwarf population identified by Gaia to
maximize the number of thick disk and halo white dwarfs, as well as
massive ($>0.8\,M_\odot$) white dwarfs in the thin disk. Massive white
dwarfs make up $\simeq10$\,\% of the population
(e.g. \citealt{tremblayetal13-1}), and a luminosity function of
$\simeq10\,000$ of these stars will determine the star formation rate
of short-lived A/B stars to within $\simeq5$\,\% throughout the age of
the thin disk with a time resolution of a few 100\,Myr
\citep{torresetal05-1}. The necessary number of fibres is small,
$\simeq10\,\mathrm{deg}^{-2}$ (Fig.\,\ref{f:gaia_wds}), only
$\simeq1-3$\,\% of the fibres available in each DESI, WEAVE, or 4MOST
field. White dwarfs are intrinsically faint, and hence to zeroth order
isotropically distributed, and should be multiplexed into all
wide-area spectroscopic surveys to maximise the total survey area /
number of observed targets.

The bulk of the Gaia white dwarf population will have $15<V<20$. Given
the proximity of most white dwarfs, even at the faint end, the Gaia
parallaxes will still be spectacularly good (5\,\% at 100\,pc for the
faintest white dwarfs in the sample). Based on our experience with
SDSS (2.5m aperture), a MOS on a 4\,m aperture should deliver a
signal-to-noise ratio of $\simeq20-30$ in a typical 1\,h exposure,
which is amply sufficient for the stellar parameter
determination. Also based on SDSS ($R\simeq1800$), we expect a radial
velocity uncertainty from the $H\alpha$ line cores measured from DESI,
WEAVE, or 4MOST low-res spectra to be $\simeq5-10$\,km/s, well
sufficient to determine the Galactic population membership.

The Gaia-based selection of white dwarfs relies on the second data
release (positions, proper motions, parallaxes, integrated XP
photometry, and $G$-magnitudes) expected in early 2017. Because of
their small radii ($\simeq0.01R_\odot$), white dwarfs will be extreme
outliers in the Gaia Hertzsprung-Russell diagram, and their
identification will be trivial and free of contaminants.  Should there
be a delay in this data release, we will fall back onto an efficient
multi-colour plus reduced proper motion selection that works very well
in the SDSS footprint \citep{gentile-fusilloetal15-1}. We are in the
process of extending this selection to the southern hemisphere making
use of the VST/ATLAS survey, as well as PanSTARRS 3pi and SkyMapper in
the near future. 

Higher-resolution ($R\simeq20\,000$) WEAVE and 4MOST spectroscopy of
the brightest Gaia white dwarfs ($V\lesssim16.5$) would help to increase the
radial velocity resolution obtained from H$\alpha$, and to improve the
sensitivity to metal pollution in Ca\,H/K, or Mg\,II 447\,nm
(Fig.\,\ref{f:sdss1228}).


%
\end{document}